\documentclass[a4paper,fleqn]{cas-dc}

\usepackage[numbers,sort]{natbib}
\usepackage{float}

\usepackage{footnote}
\usepackage{svg}

\def\tsc#1{\csdef{#1}{\textsc{\lowercase{#1}}\xspace}}
\tsc{WGM}
\tsc{QE}
\tsc{EP}
\tsc{PMS}
\tsc{BEC}
\tsc{DE}

\begin{document}
\let\WriteBookmarks\relax
\def\floatpagepagefraction{1}
\def\textpagefraction{.001}

\title [mode = title]{Attentive-based Multi-level Feature Fusion for Voice Disorder Diagnosis}     

\author[1]{Lipeng Shen}[type=editor,
                        auid=000,bioid=1,]
\fnmark[1]
\author[1]{Yifan Xiong}[type=editor,
                        auid=000,bioid=1,]
\fnmark[1]
\author[1]{Dongyue Guo}[type=editor,
                        auid=000,bioid=1,]

\author[2]{Wei Mo}[type=editor,
                        auid=000,bioid=1,]
\author[2]{Lingyu Yu}[type=editor,
                        auid=000,bioid=1,]
\author[2]{Hui Yang}[type=editor,
                        auid=000,bioid=1,]
\cormark[1]
\author[1]{Yi Lin}[type=editor,
                        auid=000,bioid=1,]
\cormark[1]

\cortext[1]{yilin@scu.edu.cn (Y. Lin), yh8806@163.com (H. Yang)}
\fntext[1]{Equal Contributions.}

\credit{Conceptualization of this study, Methodology, Software}

\affiliation[1]{organization={College of Computer Science, Sichuan University},
                addressline={No.24 South Section 1, Yihuan Road}, 
                city={Chengdu},
                postcode={610065}, 
                state={Sichuan},
                country={China}}
\affiliation[2]{organization={Department of Otorhinolaryngology-Head \& Neck Surgery, West China Hospital, Sichuan University},
                addressline={No.24 South Section 1, Yihuan Road}, 
                city={Chengdu},
                postcode={610065}, 
                state={Sichuan},
                country={China}}

\begin{abstract}
Voice disorders negatively impact the quality of daily life in various ways. However, accurately recognizing the category of pathological features from raw audio remains a considerable challenge due to the limited dataset. A promising method to handle this issue is extracting multi-level pathological information from speech in a comprehensive manner by fusing features in the latent space. In this paper, a novel framework is designed to explore the way of high-quality feature fusion for effective and generalized detection performance. Specifically, the proposed model follows a two-stage training paradigm: (1) ECAPA-TDNN and Wav2vec 2.0 which have shown remarkable effectiveness in various domains are employed to learn the universal pathological information from raw audio; (2) An attentive fusion module is dedicatedly designed to establish the interaction between pathological features projected by EcapTdnn and Wav2vec 2.0 respectively and guide the multi-layer fusion, the entire model is jointly fine-tuned from pre-trained features by the automatic voice pathology detection task. Finally, comprehensive experiments on the FEMH and SVD datasets demonstrate that the proposed framework outperforms the competitive baselines, and achieves the accuracy of 90.51\% and 87.68\%.
\end{abstract}


\maketitle

\section{Introduction}
Voice disorders are extensively acknowledged as one of the most serious diseases by disrupting communication and affecting the quality of physical, social, and professional life \cite{hegde2019survey}. Approximately 25\% of the general population suffers from voice disorders, especially in professions that excessively rely on voice \cite{al2017investigation}. However, the traditional diagnostic method, namely laryngoscopy, is relatively expensive and uncomfortable due to the invasive examination, which leads patients to avoid it and miss the optimal treatment timing \cite{fang2019detection}, thereby leading to the negative progression of the disorder.

Considering the abovementioned issue, automatic voice pathology detection (AVPD) as a non-invasive ‌reliable voice diagnosis tool attracts widespread attention, which extracts task-oriented information to assign a predefined pathological category to each utterance by machine learning-based techniques \cite{amara2016improved}. In general, the AVPD framework consists of the feature extraction module and detection module. In the former, current related works demonstrate that traditional acoustic features, such as MFCC \cite{tirronen2022effect} and Mel Spectrogram \cite{javanmardi2024comparison}, have the ability to initially extract task-oriented information from raw audio. However, traditional features are constrained by the need for manually configured hyperparameters, which may lead to the leakage of pathological information.


Furthermore, machine learning-based classifiers, i.e., SVM \cite{hammami2016pathological}, GMM \cite{amara2016improved}, RF \cite{yuanbo2020voice}, and KNN \cite{chen2021voice}, face the problem of poor feature matching and computational performance in high-dimensional spaces. Therefore, the following works focused on the deep learning-based methods that show effective performance across various domains, and deep learning-based detection module is becoming the mainstream classifier in AVPD task  \cite{harar2017voice, alhussein2018voice}. Due to the excellent ability of neural networks to capture pathological information, these methods preliminarily solve the problems caused by machine learning-based classifiers and achieve the desired performance. 

However, deep learning-based pathology detection methods are still facing the following limitations: 
1) Due to ethical issues, the size of training samples is generally limited, thereby obtaining undesirable performance and generalizability of the data-driven methods.
2) Considering the limited training samples, it is a challenging task to balance the model parameter size for maximizing the extraction of disorder-related information from pathological utterances.

To address these problems mentioned above, in this paper, we propose a novel feature fusion-based disorder detection framework to effectively improve AVPD performance by a multi-stage training strategy. Due to the limited datasets, while adopting ECAPA-TDNN \cite{desplanques2020ecapa} to extract the task-oriented feature from MFCC, we also employ the Wav2vec 2.0 \cite{baevski2020wav2vec} to obtain acoustic representation as supplementary to enhance the pathological information. Subsequently, the attention mechanism is used to guide the temporal-fusion operation between the output embeddings of separate models to obtain robust embedding, which benefits achieving desired performance on limited datasets. Considering the increasing trainable parameters, instead of end-to-end training, we fine-tune the proposed model with a multi-stage strategy, which allows the model to integrate prior pathological knowledge into AVPD in a curriculum manner. 

To validate the final performance, we conduct experiments on open-source FEMH and SVD datasets. The results demonstrate that the proposed model outperforms selective baselines, and achieves Accuracy of 90.51\% and 87.68\% respectively. The proposed modules contribute desired performance improvements and the visualization also validates the effectiveness of the proposed model. It is believed that the proposed method can provide a promising solution for the AVPD task.

\section{Related work}
In recent decades, extensive works have focused on AVPD tasks, which can be divided into machine learning-based and deep learning-based methods.

\subsection{Machine leanring-based methods}

Due to the ability to capture the underlying correlations among inputs, machine learning is widely employed in the field of AVPD tasks. Al-Dhief et al. \cite{al2021voice} employ the OSELM algorithm as the disorder voice classifier to detect the common three types of pathologies (cyst, polyp, and paralysis). Fan et al. \cite{fan2020modeling} adopt three imbalanced learning algorithms, i.e., SMOTE, Borderline-SMOTE, and ADASYN, to oversample the minority class, which alleviates inadequate accuracy for long-tail pathology voice datasets but reduces the head-class accuracy in return. A FLN-based method is designed in \cite{albadr2024fast}, and the results are evaluated by a range of objective metrics. Tirronen et al. \cite{tirronen2022effect} choose the most widely used classifier, namely SVM, to study the effect of the MFCC frame length in AVPD on the SVD subset. In summary, the machine learning-based methods achieved some progress in improving disorder detection performance. However, the inherent limitations of machine learning impact the model performance, resulting in poor generalization and failure to handle complex tasks.

\subsection{Deep leanring-based methods}

Considering the limitations of machine learning, deep learning-based models, demonstrating effectiveness across various domains, are gradually applied in AVPD tasks as alternatives and make certain achievements. In \cite{mohammed2020voice}, the pre-trained convolutional neural network was applied to the SVD dataset, which improve the classification accuracy over machine learning-based methods. Ksibi et al. \cite{ksibi2023voice} cascaded CNN and RNN to judge the gender to support the decision of whether the voice is pathological or healthy. Jegan et al. \cite{jegan2023voice} constructed CNN-based networks to capture inherent speech variations, which is beneficial for pathological voice detection from MFCC, and employed the artificial bee colony optimization algorithm to fine-tune the proposed network. Although the model performance can be improved, existing naive deep learning methods still limit the model's capacity to capture task-related information from pathological voice datasets. To address this issue, many methods were tried to fully extract the robust feature from voice, among which the fusion-based method is widely adopted. Ankışhan et al. \cite{ankicshan2021voice} believed that utilizing available data in a complementary manner towards the learning of a complex task can improve the model performance. Two fusion strategies, namely feature-level and decision-level fusion, were applied to improve the classification accuracy of the multi-model mode. Mohammed et al. \cite{mohammed2023mmhfnet} employed the skip connect mechanism to capture multi-layer information, and concatenate them for effective fusion. In summary, while it can be seen that current fusion methods relatively alleviate the issue of limited data compared with naive neural networks, only simple fusion approaches still raise concerns about the reliability of model performance.

\section{Methods}
\subsection{Overall framework}

\begin{figure*}[!t]
	\centering
	\includegraphics[width=.9\textwidth]{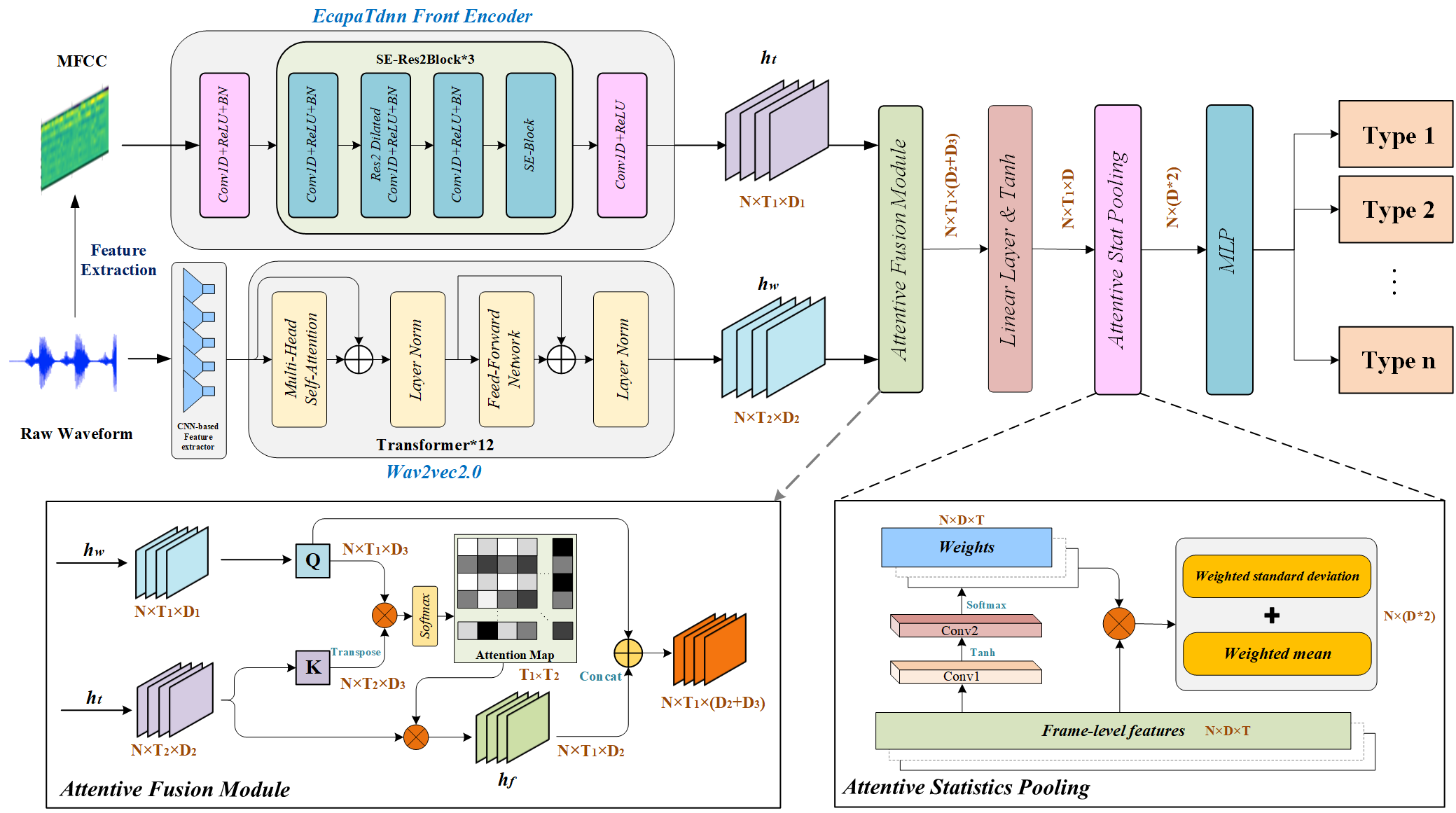}
	\caption{The architecture of proposed framework}
	\label{FIG:1}
\end{figure*}

As illustrated in Fig.\ref{FIG:1}, ECAPA-TDNN serves as the backbone network, which is enhanced by the pre-trianed Wav2vec 2.0. Specifically, the proposed framework is composed of a feature extraction module, an attentive fusion module, and a disorder detection module. The feature extraction module, including the front encoder of ECAPA-TDNN and Wav2vec 2.0, is adopted to extract high-dimensional embeddings from MFCC and raw waveform $x$, respectively. Subsequently, the attentive fusion module achieves temporal fusion to project the multi-level pathological information into the robust feature, based on which the disorder detection module performs the AVPD task using a classification head. The detailed implementation of each module is described as follows.

\subsection{Feature extraction module}

In general, the handcrafted features, i.e., MFCC and Melspectrogram, are used for speech-related tasks, which may not be optimal to represent the voice pathological information. In addition, with the limited dataset volume, comprehensive information extraction is crucial for subsequent classification processing. To address this issue, while using the front encoder of ECAPA-TDNN to extract the embedding $h_t$ from MFCC, we also employ the Wav2vec 2.0 with pre-trained 960-hour Librispeech \cite{panayotov2015librispeech} to obtain the embedding $h_w$ in a parallel manner as the supplement to $h_t$. The details of the dual-path feature extraction is illustrated as follows:

\textbf{ECAPA-TDNN path} 
This path aims to capture the disorder-related information and adjust the embedding dimension for the subsequent processing. We define the ECAPA-TDNN prior to the ASP as the front encoder to complete this task. Specifically, the front encoder of ECAPA-TDNN cascades two Conv1D layers and three SE-ResBlocks to extract the high-dimensional speech representation ${h_t} = \left[ {h_t^1,h_t^2, \cdots,h_t^n} \right]$ from MFCC.

\textbf{Wav2vec 2.0 path}
Due to the limitation of MFCC in extracting the disorder-related information from pathological utterances, Wav2vec 2.0 is employed as the complement to the ECAPA-TDNN path, consisting of a convolution-based block and 12 Transformer blocks. To obtain the effective representation ${h_w} = \left[ {h_w^1,h_w^2, \cdots ,h_w^n} \right]$ with limited training samples, Wav2vec 2.0 pretrains with the 960-hour Librispeech, and is fine-tuned by AVPD task.


\subsection{Attentive fusion module}
Learning robust task-oriented embeddings from the universal acoustic representation is effective in improving the model performance. Moreover, it is believed that various frame of separate representations makes unique contributions to the fusion process. Considering the efficiency and effectiveness of the attention mechanism in other tasks \cite{guo2023boosting, hu2018squeeze, vaswani2017attention}, the attention mechanism is applied to guide the fusion between $h_t$ and $h_w$ for obtaining the robust disorder-related information. Specifically, the attentive fusion module comprises two blocks, i.e., ECAPA-TDNN query (Q) block and Wav2vec 2.0 key (K) block, which project $h_t$ and $h_w$ into the fusion space separately as implemented by the FC layer. Mathematically, the calculation process can be formulated as follows:
\begin{equation}
    q = {h_t}{W_q} + b_q
\end{equation}
\begin{equation}
    k = {h_w}{W_k} + b_k
\end{equation}
where the $W_q$ and $W_k$ are the learnable weight, and $b_q$ and $b_k$ are the bias. The $q \in {R^{T \times D}}$ and $k \in {R^{T \times D}}$, D represents the dimension of representations, and $T$ denotes the temporal dimension.

Subsequently, the frame-wise probability scores are generated by $q$ and $k$ to represent the importance weights of each frame and fuse them with $h_t$ to obtain the pathology-enhanced embedding $h_f$. To achieve task-related information maximization during feature fusion, representation $h_w$ output by Wav2vec 2.0 is concatenated with $h_f$ to generate the final fused embedding $h_r$, as shown below:
\begin{equation}
    Scores = softmax(\frac{{Q{K^T}}}{{\sqrt {{d_k}} }})
\end{equation}
\begin{equation}
    {h_r} = concat[Scores \times {h_t},{h_w}] 
\end{equation}
where $d_k$ is the scaling factor.

\subsection{Multi-stage training strategy}
In general, the speech classification models are trained by the end-to-end strategy, which can avoid error accumulation and enhance performance by directly optimizing the final target loss during the training stage. However, due to the considerable model capacity of the feature extraction module, this end-to-end strategy may not be optimal in this work. 

To this end, we propose a multi-stage training strategy to optimize the model in a curriculum way. In practice, we train the ECAPA-TDNN $\varphi $ and Wav2vec 2.0 $\phi $ paths to implement AVPD task respectively to learn the primary pathological information in the first stage:
\begin{equation}
    {\mathop {\min }\limits_\varphi  CE(MLP(Avgpool({h_t})),label)}
\end{equation}
\begin{equation}
    {\mathop {\min }\limits_\phi  CE(MLP(Avgpool({h_w}),label)}
\end{equation}
where $CE$ denotes the cross-entropy loss, $Avgpool$ refers to the global average pooling and $label$ indicates the specific disorder type associated with each audio recording.

Subsequently, based on the first training stage, we reload the pre-trained weights and fine-tune the whole model $\theta $ in the second stage, which enables the model to optimize the attentive fusion module to achieve performance improvement. The optimization process is shown as follows:
\begin{equation}
    \mathop {\min }\limits_\theta  CE(logit,label)
\end{equation}
where $logit$ represents output of the entire model which indicates the predicted likelihood of the disorder.

\section{Experiments}
\subsection{Data descriptions}
To validate the effectiveness and generalizability of the proposed model, we conduct experiments on two open-source datasets, i.e., the Far Eastern Memorial Hospital Language Disorder (FEMH) dataset \cite{femhdataset} and the Saarbrücken Voice Disorder (SVD) dataset \cite{woldert2007saarbruecken}. 

\textbf{FEMH}: All speech recordings were collected by Far Eastern Memorial Hospital's Speech Clinic from 2012 to 2019, which can be classified into four categories of disorders: phonotrauma (Pho), vocal palsy(Vp), functional dysphonia(FD), and neoplasm(Neo). The waveforms were recorded at a sampling rate of 44.1 kHz with 16-bit resolution via high-quality microphones and digital amplifiers, with background noise levels ranging from 40 to 45 dBA. The dataset includes vowels /a/, /i/, /u/, and sentence recordings. Following previous work \cite{mohammed2023mmhfnet}, we choose the language-independent vowel /a/ and sentence subsets to conduct the following experiments.

\textbf{SVD}: The SVD dataset was developed by the Institute of Phonetics from the University of Saarland, containing vowels and sentences for speech and EGG modalities. The vowel recordings consist of /a/, /i/, /u/ in normal, high, low and low-high-low intonation. The sentence recordings are the sentence "Good morning, how are you?". The speakers range from 16 to 80 years old, and the speech is sampled at 50 kHz with 16-bit resolution. In this work, we categorize all recordings into pathology and health classes to implement the AVPD task.

\subsection{Implementation details}
Considering the sample rate of the Librispeech dataset is 16kHz, the raw audios are downsampled to 16 kHz for both FEMH and SVD datasets before feeding into the Wav2vec 2.0 path of the feature extraction module. During the training process, the Adam optimizer is employed to optimize the model and cross-entropy serves as the loss function. Specifically, in the first training stage, the learning rate is 1e-5 for the front encoder of ECAPA-TDNN and 1e-4 for Wav2vec 2.0. In the second training stage, the learning rate is 1e-5.
The batch size is set to 64. 
The proposed model is implemented using Pytorch 2.2.2 and all experiments are conducted on two NVIDIA GeForce RTX 3090 with 48GB memory.

\subsection{Baselines and Evaluation metrics}
\textbf{Baselines}: To demonstrate the effectiveness of the proposed framework, we select three state-of-the-art models, namely SincNet \cite{hung2022using}, CSILNet \cite{wang2022continuous} and SMMFNet \cite{mohammed2023mmhfnet}, to evaluate the performance. \textbf{SincNet} designs a signal processor with a series of learnable Sinc filters to extract the acoustic information from waveforms before the CNN/DNN layers. 
Most importantly, the Sinc function-based framework shows the expected performance with sufficient interpretability. \textbf{CSILNet} implements AVPD tasks using sentence-based speech signals. This framework converts the audio signals into MFCC, and then applies BiLSTM to capture the sequential relationships within the sentences, aiming to extract disorder-related information from the sentences. 
\textbf{SMMFNet} utilizes multi-level features extracted from different layers of the CNN, which are fused through concatenation. Subsequently, the LSTM-based classifier is employed to capture the information across these layers, thereby effectively enhancing performance. 

\textbf{Evaluation metrics}: In order to validate the performance of the proposed model, a total of four evaluation metrics are considered from other classification tasks \cite{guo2023comparative}, which include accuracy (Acc), precision (Pre), recall (Rec), and F1 score (F1), to quantitatively evaluate the model performance. 
For all four metrics, a larger value indicates higher performance.
\begin{table}[width=.9\linewidth, pos=h]
\renewcommand{\arraystretch}{2}
\setlength{\tabcolsep}{3pt}
\scriptsize
\caption{Comparsion with competitive baselines on the FEMH and SVD} 
\resizebox{\linewidth}{!}{ 
\label{tab:t4}    
\centering
\begin{tabular}{ccccccc}
\hline
\textbf{Dataset}      & \textbf{Type}          & \textbf{Method} & \textbf{Acc \%} & \textbf{F1 \%} & \textbf{Pre \%} & \textbf{Rec \%} \\ \hline
\multirow{4}{*}{FEMH} & \multirow{2}{*}{Vowel} & SincNet  & 72.18 & - & -     & -  \\
                      &                        & Proposed & 85.52        & 85.41       & 85.97              & 85.52  \\ \cline{2-7}
                      & \multirow{2}{*}{Sentence}   & CSILNet  & 89.27 & - & -     & -  \\
                      &                        & Proposed & 90.51 & 90.44 & 91.48     & 90.51  \\ \hline                  
\multirow{4}{*}{SVD}  & \multirow{2}{*}{Vowel} & SMMFNet  & 78.19 & 84.47 & 81.21     & 88.00  \\
                      &                        & Proposed & 79.17        & 83.58       & 79.71              & 87.84  \\ \cline{2-7}
                      & \multirow{2}{*}{Sentence}   & SMMFNet  & 85.43 & 89.61 & 87.11     & 92.25  \\
                      &                        & Proposed & 87.68 & 91.30 & 93.12     & 89.55  \\ \hline
\end{tabular}
}
\end{table}

\subsection{Comparison with the state-of-the-art}
The experimental results of the FEMH and SVD datasets are reported in Table \ref{tab:t4} in terms of the proposed metrics. Considering that existing baselines are only measured by Acc, in this work, the Acc serves as the primary metric to evaluate the model performance. As shown in Table \ref{tab:t4}, the proposed model achieves the highest accuracy of 85.52\% (vowel) and 90.51\% (sentence) for FEMH, while 79.17\% (vowel) and 87.68\% (sentence) for SVD dataset, which outperforms all selected baselines. 
Specifically, the following key points can be concluded:

(a) Thanks to the combination of attentive fusion and training strategy, the proposed method achieves improvements of 13.34\% and 1.24\% for vowel and sentence subsets of the FEMH dataset over the best baselines, while enhancing performance by 0.98\% and 2.25\% for vowel and sentence subsets of the SVD dataset. The results demonstrate that incorporating the abovementioned contributions has been shown to significantly enhance model performance on the AVPD task, further validating the motivation behind our work.

(b) Compared to the vowels, on both the FEMH or SVD datasets, the proposed model shows superior performance on the sentence subset. This finding demonstrates the effectiveness of the model in extracting pathological information from sentences. Moreover, it is evident that sentence speeches can provide more pathological information than that of in vowel speeches. This observation can also be supported by our general understanding, i.e., sentence speeches with richer contextual information are preferred as the input of AVPD tasks.

(c) Among the four datasets in Table \ref{tab:t4}, the proposed model achieves the most significant improvement on the shortest vowel recordings from the FEMH dataset (containing the least pathological information).
The results can be attributed that the multi-level feature learning and the fusion-based method in this work can effectively enhance the learning ability of the pathological information to support the AVPD tasks.

(d) Although the proposed model outperforms selective baselines in terms of selected metrics on most datasets, it only obtains limited performance improvement on the SVD-vowel dataset, i.e., with higher accuracy
and inferior F1 score, recall, and precision.

It is believed that the diverse pathological categories (71 disorders)  in the SVD-vowel dataset provide challenges to learning task-oriented features to formulate preferred classification.


In summary, the experimental results demonstrate that the proposed model harvests the expected performance on both FEMH and SVD, and also shows the effectiveness of the abovementioned contributions.
\begin{table}[width=.9\linewidth, pos=h]
\renewcommand{\arraystretch}{2}
\setlength{\tabcolsep}{3pt}
\scriptsize
\caption{Comparsion with single path and fusion model on the FEMH and SVD} 
\label{tab:t1}    
\centering
\begin{tabular}{ccccccc}
\hline
\textbf{Dataset}      & \textbf{Type}          & \textbf{Method} & \textbf{Acc \%} & \textbf{F1 \%} & \textbf{Pre \%} & \textbf{Rec \%} \\ \hline
\multirow{6}{*}{FEMH} & \multirow{3}{*}{Vowel} & ECAPA-TDNN  & 65.22 & 65.82 & 66.69     & 65.22  \\
                      &                        & Wav2vec2.0 & 79.71 & 79.16 & 81.44     & 79.71  \\
                      &                        & Fusion & 85.52        & 85.41       & 85.97              & 85.52  \\ \cline{2-7}
                      & \multirow{3}{*}{Sentence}   & ECAPA-TDNN  & 87.32 & 86.97 & 87.03     & 87.32  \\
                      &                        & Wav2vec2.0 & 85.71 & 85.17 & 85.74     & 85.71  \\
                      &                        & Fusion & 90.51 & 90.44 & 91.48     & 90.51  \\ \hline
\multirow{6}{*}{SVD}  & \multirow{3}{*}{Vowel} & ECAPA-TDNN  & 73.28 & 71.96 & 72.15     & 73.28  \\
                      &                        & Wav2vec2.0 & 77.45 & 77.37 & 77.30     & 77.45  \\
                      &                        & Fusion & 79.17        & 83.58       & 79.71              & 87.84  \\ \cline{2-7}
                      & \multirow{3}{*}{Sentence}   & ECAPA-TDNN  & 84.17 & 83.71 & 83.82     & 84.17  \\
                      &                        & Wav2vec2.0 & 85.43 & 84.73 & 85.44     & 85.43  \\
                      &                        & Fusion & 87.68 & 91.30 & 93.12     & 89.55  \\ \hline
\end{tabular}
\end{table}

\subsection{Ablation studies}
\textbf{Effectiveness of feature fusion}: To determine the effectiveness of proposed attention feature fusion in extracting the robust representation, we use both single models, i.e., ECAPA-TDNN and Wav2vec 2.0, and the fusion model to implement the AVPD task according to four selected metrics. Unlike ECAPA-TDNN, Wav2vec 2.0 is commonly used to extract the temporal embedding. Thus, in practice, we conduct the experiments with Wav2vec 2.0 in the following ways: loading the pre-trained weights of Wav2vec 2.0 on the 960-hour LibriSpeech, and we cascade the Wav2vec 2.0 and Multi-layer Perceptron (MLP) which contains three linear layers (with neurons 768, 128, 4 for FEMH or 2 for SVD). In this context, the global average pooling is employed to compress the temporal dimension before MLP. 

The results are listed in Table \ref{tab:t1}. It is clear that the proposed fusion model outperforms its counterparts across all objective evaluation metrics on both the vowel and sentence datasets. The results indicate that, compared to the individual models (ECAPA-TDNN and Wav2vec 2.0), the fused embedding $h_r$ is more effective in capturing task-specific information, which is essential for improving model accuracy and robustness in downstream tasks. Moreover, these findings further support the motivation behind the feature fusion strategy.
\begin{table}[pos=h]
\renewcommand{\arraystretch}{2}
\caption{Comparision with various fusion styles on the FEMH and SVD} 
\resizebox{\linewidth}{!}{ 
\label{tab:t2}    
\centering
\begin{tabular}{cccccc}
\hline
\textbf{Subset}     & \textbf{Methods} & \textbf{Acc \%} & \textbf{F1 \%} & \textbf{Pre \%} & \textbf{Rec \%} \\ \hline
\multirow{3}{*}{FEMH-vow} & Add              & 84.06        & 84.38       & 85.32              & 84.06           \\
                   & Concat           & 84.06       & 83.90       & 84.43              & 84.06           \\
                   & Attention        & 85.52        & 85.41       & 85.97              & 85.52           \\ \hline

\multirow{3}{*}{FEMH-sen} & Add              & 88.23        &  89.04       & 89.11              & 88.23           \\
                   & Concat           & 88.71        & 87.06       & 87.85              & 88.71           \\
                   & Attention        & 90.51        & 90.44       & 91.48              & 90.51           \\ \hline
\multirow{3}{*}{SVD-vow} & Add              & 77.62        & 80.25       & 78.11              & 82.51           \\
                   & Concat           & 78.29       & 81.62       & 78.83              & 84.62           \\
                   & Attention        & 79.17        & 83.58       & 79.71              & 87.84           \\ \hline
\multirow{3}{*}{SVD-sen} & Add              & 85.84        &  87.50       & 87.23              & 87.77           \\
                   & Concat           & 86.47        & 89.00       & 89.44              & 88.57           \\
                   & Attention        & 87.68        & 91.30       & 93.12              & 89.55           \\ \hline
                   
\end{tabular}
}
\end{table}

\textbf{Seclection of fusion method}: As described before, feature fusion is effective on the AVPD task. However, except the proposed attentive fusion in this paper, other feature fusion modules are also widely applied in deep learning models\cite{mohammed2023mmhfnet, tariq2022feature}, such as Add, and Concat, which significantly reduces the number of trainable parameters compared with the attentive fusion. 
To confirm the effectiveness of the proposed fusion, FEMH and SVD are selected to conduct ablation experiments, and the results are reported in Table \ref{tab:t2}. It is evident that the attentive fusion consistently outperforms the other methods across all selected metrics, based on both the vowel and sentence dataset. This demonstrates that the proposed attentive fusion method enables the model to dynamically fuse the $h_t$ and $h_w$ in the latent space. Furthermore, the superior performance of attentive fusion compared to other fusion strategies also highlights its effectiveness in producing more informative and task-relevant embeddings. In conclusion, attentive fusion proves to be the most effective fusion approach among selected fusion strategies.


\begin{figure*}[htbp]
    \centering
    \includegraphics[width=0.3\textwidth]{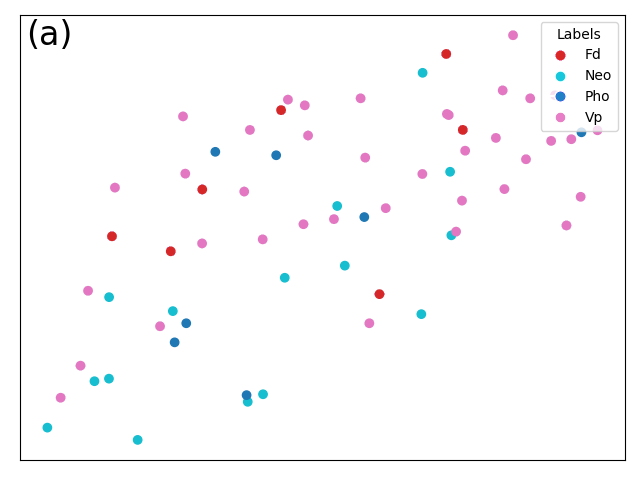}
    \includegraphics[width=0.3\textwidth]{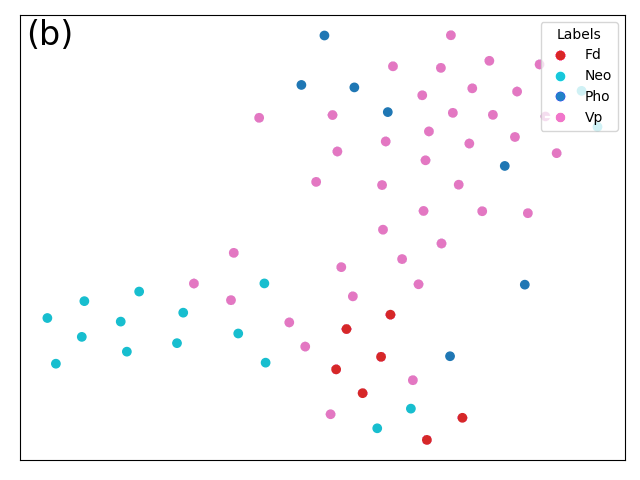}
    \includegraphics[width=0.3\textwidth]{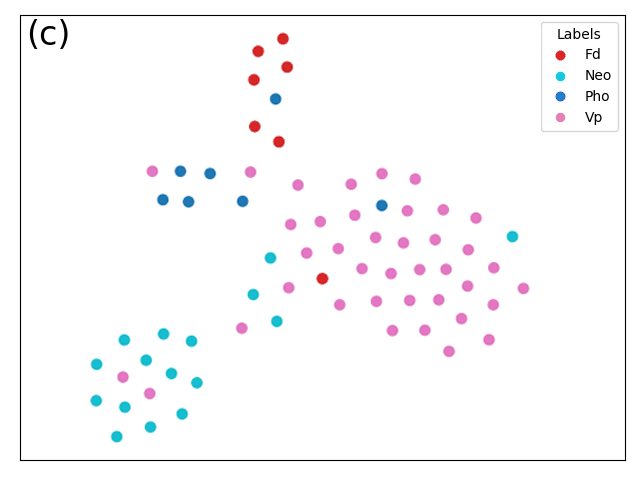}
    \caption{Visualization of embeddings projected by (a) ECAPA-TDNN, (b) Wav2vec 2.0, and (c) proposed model}
    \label{fig:cluster}
\end{figure*}

\begin{table}[pos=h]
\renewcommand{\arraystretch}{2}
\caption{Comparision with training strategies on the FEMH and SVD} 
\resizebox{\linewidth}{!}{ 
\label{tab:t3}    
\centering
\begin{tabular}{cccccc}
\hline
\textbf{Subset}     & \textbf{Strategy} & \textbf{Acc \%} & \textbf{F1 \%} & \textbf{Pre \%} & \textbf{Rec \%} \\ \hline
\multirow{2}{*}{FEMH-vow} & end-to-end     & 81.24        & 81.11       & 82.29              & 81.24           \\
                   & multi-stage           & 85.52       & 85.41       & 85.97              & 85.52           \\ \hline

\multirow{2}{*}{FEMH-sen} & end-to-end              & 87.54        &  87.66       & 87.31              & 87.54           \\
                   & multi-stage        & 90.51        & 90.44       & 91.48              & 90.51           \\ \hline

\multirow{2}{*}{SVD-vow} & end-to-end     & 78.23        & 79.80       & 78.41              & 81.24           \\
                   & multi-stage           & 79.17       & 83.58       & 79.71              & 87.84           \\ \hline

\multirow{2}{*}{SVD-sen} & end-to-end              & 86.04        &  87.66       & 88.19              & 87.13           \\
                   & multi-stage        & 87.68        & 91.30       & 93.12              & 89.55           \\ \hline
\end{tabular}
}
\end{table}

\textbf{Results of multi-stage training strategy}:
To further explore the effectiveness of the multi-stage training strategy, ablation experiments on the FEMH and SVD datasets are presented in Table \ref{tab:t3}.

As described above, the multi-stage training strategy can enhance the capability of the attentive fusion module, and obtain $h_r$ with robust pathological information compared to the end-to-end strategy. In this section, the following steps are applied in experiments: 1) we train the proposed model from scratch with a cross-entropy loss function, i.e., the end-to-end training strategy. 2) Firstly, we train the front encoder of ECAPA-TDNN and Wav2vec 2.0 as the pre-trained feature extractors and subsequently use the cross-entropy to fine-tune the entire model. The results are reported in \ref{tab:t3}, and it can be found that on both the FEMH and SVD datasets, the multi-stage training harvests expected performance improvements, indicating that the fused embedding trained by multi-stage outperforms the end-to-end strategy. To be specific, compared with the end-to-end training strategy, the model performance trained by multi-stage is improved by 4.28\% (vowel) and 2.97 \% (sentence) on the FEMH dataset, while the improvement reaches 0.94\% (vowel) and 1.64\% (sentence) on the SVD dataset respectively.

\subsection{Visualization}

To visually demonstrate the performance of the proposed model, we employ the t-SNE algorithm to illustrate the classification embeddings of the proposed model and Wav2vec 2.0. Specifically, the high-dimensional embeddings before the final classification linear layer are selected to formulate the cluster in the latent space.

As detailed in Fig.\ref{fig:cluster} (a) and (b), we can obviously see that the embeddings of different categories are evidently coupled, failing to form distinct feature clusters, which means the ECAPA-TDNN and Wav2vec 2.0 fail to provide the effective embedding for the classification task. Fortunately, the features projected by the proposed model generate compact intra-class clusters and independent inter-class boundaries, indicating that the proposed model has the capability to extract the task-oriented feature with the limited dataset, as shown in Fig.\ref{fig:cluster} (c).

\section{Conclusions}
In this work, we propose a novel fusion-based framework to achieve the AVPD task with limited training samples, which harvests a relatively high performance. Specifically, an attentive fusion module is designed to enhance the pathological information in fused feature, and the entire model training progress can be divided into the following two steps, i.e., extracting the primary pathological information by ECAPA-TDNN and Wav2vec 2.0 respectively, and fine-tuning the whole model with pre-trained embeddings. Experiments are conducted across the two open-source datasets in terms of four metrics. The experimental results demonstrate that the attentive fusion module significantly improves the AVPD performance compared with ECAPA-TDNN and Wav2vec 2.0. Furthermore, thanks to the multi-stage training strategy, the proposed model achieves further performance improvements.
Most importantly,  the proposed model obtains the best accuracy of 90.51\% and 87.68\% respectively with limited data samples, which is also qualitatively validated by the visualization cluster.

\bibliographystyle{cas-model2-names}
\bibliography{ref}

\end{document}